# DFT-Spread OFDM with Frequency Domain Reference Symbols

Alphan Şahin, Erdem Bala, Rui Yang, Robert L. Olesen
{Alphan.Sahin, Erdem.Bala, Rui.Yang, Robert.Olesen}@InterDigital.com

*Abstract*—The fifth generation (5G) wireless standard will support several new use cases and 10 to 100 times the performance of fourth generation (4G) systems. Because of the diverse applications for 5G, flexible solutions which can address conflicting requirements will be needed. In this paper, we propose a solution which enables the use of discrete Fourier transform spread orthogonal frequency division multiplexing (DFT-s-OFDM) and OFDM, which address different requirements, using a common reference symbol (RS) design. In this solution, the DFT-s-OFDM symbol contains RSs in the frequency domain that may be shared by a subsequent OFDM symbol. The proposed scheme is generated by puncturing the output of a DFT-spread block and replacing the punctured samples with RSs in frequency. We prove that puncturing the interleaved samples at the output of the DFT-spread operation equivalently introduces a periodic interference to the data symbols at the input of the DFT-spread operation. We show that the interference due to the puncturing can be removed with a low-complexity receiver by exploiting the zeros inserted to certain locations before the DFT-spread block at the transmitter. Simulation results support that the proposed scheme removes the error floor caused by the puncturing and achieves lower peak-to-average-power ratio than OFDM.

## I. INTRODUCTION

To support diverge applications and wide range of deployment scenarios in future wireless communication systems, the 5G standard will provide flexible features that can efficiently utilize the resources in a given situation. One of the most important features considered in 5G standard is to use different waveforms such as orthogonal frequency division multiplexing (OFDM) and DFT-spread OFDM (DFT-s-OFDM), depending on different channel conditions and deployment scenarios in the uplink [1]. To facilitate this essential feature, it is highly desirable to use a flexible reference signal (RS) design, in which the RS allocation in time and frequency is common for different waveforms. Common RS allocation will be beneficial for the system as the receiver can estimate the channel regardless of the waveform type that is used. A reliable channel estimation can therefore be achieved in coexistence scenarios such as uplink multi-user multi-input, multi-output (MIMO) cases.

In basic DFT-s-OFDM, the data symbols are first spread with a DFT block, and then mapped to the input of an IDFT block. On the other hand, OFDM does not employ a DFT-spread block and the symbols are mapped to the subcarriers directly. Hence, while the RSs are distributed in frequency for OFDM, they are multiplexed in time for DFT-s-OFDM. Therefore, the coexistence of DFT-s-OFDM and OFDM leads to inconsistent overlapping on the same resources. One method to obtain a common framework for the reference symbols is to dedicate a separate symbol for RSs, which increases the overhead as mentioned earlier. Another method is to introduce RSs in frequency domain for different waveforms. Although this option is straightforward for OFDM, it is not trivial for DFT-s-OFDM as the data symbols are spread across the subcarriers with this waveform. In this study, we will address this challenge.

In the literature, there are notable studies on reference symbols for DFT-s-OFDM and OFDM [2-7]. For example, in [3], data symbols and RSs are multiplexed in the same OFDM symbol while minimizing the PAPR of the signals. Although this method has its own merit, it expands the number of resources utilized in frequency, which may not be desirable as it increases the system complexity in multi-user scenarios. In [4], the RSs are added to the two edges of the inputs of the DFT-s-OFDM, which substitute the function of cyclic prefix (CP). In [5], it has been shown that phase noise can be compensated by inserting uniformly distributed RSs before DFT-spread block. In LTE, demodulation (precoded) RSs are inserted into some of the OFDM symbols in a subframe in the downlink [2]. LTE also has a, channel state information (CSI) RS, which is an unprecoded RS employed to estimate the downlink channel state information [6], generated by puncturing the data symbols in frequency for OFDM symbols. However, UEs that are LTE release 8/9 compliant are not aware of the existence of the CSI-RS, and will demodulate the CSI-RS as interference. A puncturing operation in frequency for DFT-s-OFDM is also mentioned in [7] with the motivation of adjusting the coding rate in frequency. To the best of our knowledge, the systematic analysis of the transmitter and receiver structures of punctured DFT-s-OFDM for the sake of achieving RSs in frequency domain is not available in the literature.

In this study, we show that it is possible to synthesize punctured DFT-s-OFDM symbols that can be demodulated with a low-complexity receiver. We prove that if the interleaved samples at the output of a DFT operation are punctured, a periodic interference is introduced into the original symbols at the input of the DFT. By exploiting the periodic interference structure, we show that inserting zeros to certain locations before the DFT-spread block at the transmitter allows the receiver to estimate the impact of distortion due to the puncturing and enables the cancelation of the interference. In addition, we investigate two receiver structures, i.e., low complexity receiver and iterative receiver. We compare the performance of these receivers with a plain DFT-s-OFDM design.

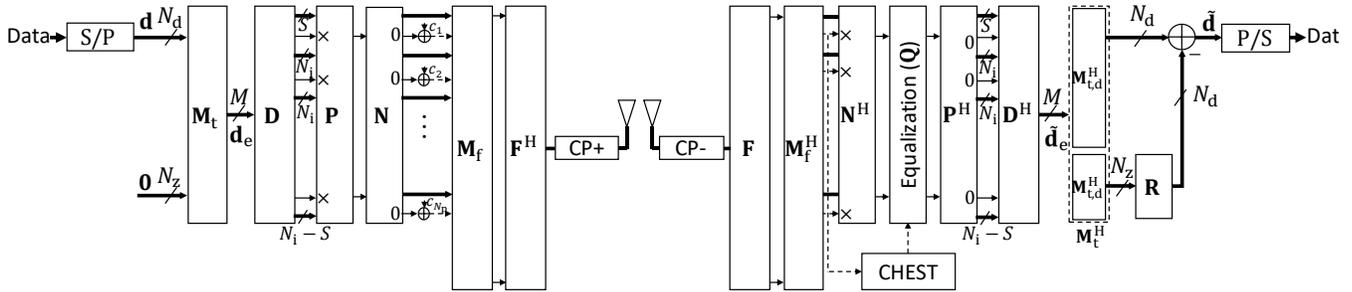

Fig. 1. Transmitter and receiver structures for punctured DFT-s-OFDM with frequency domain reference symbols.

The rest of the paper is organized as follows. In Section II, we provide the system model for DFT-s-OFDM. In Section III, we discuss the system model for punctured DFT-s-OFDM, and introduce the receiver structures which exploit a periodic interference due to an interleaved puncturing. In addition, we discuss the limitations and a generalization of the proposed scheme in this section. In Section IV, we compare the proposed scheme and plain DFT-s-OFDM. Finally, we conclude the paper with closing remarks in Section V.

Notations: Matrices [columns vectors] are denoted with upper [lower] case boldface letters (e.g., $\mathbf{A}$ and [$\mathbf{a}$]). The Hermitian operation and the transpose operation are denoted by $(\cdot)^{\mathrm{H}}$ and $(\cdot)^{\mathrm{T}}$, respectively. The symbol $\otimes$ denotes Kronecker product. The field of complex numbers, the field of real numbers, and the field of natural numbers are shown as $\mathbb{C}$, $\mathbb{R}$, and $\mathbb{N}$, respectively. $\mathbb{N}_0$ denotes $\mathbb{N} \cap \{0\}$. The multivariate complex Gaussian distribution is denoted by $\mathcal{CN}(\boldsymbol{\mu}, \mathbf{C})$, where $\boldsymbol{\mu}$ is the mean vector and $\mathbf{C}$ is the covariance matrix. $\mathbf{I}_N$, $\mathbf{0}_{N\times M}$, $\mathbf{1}_{N\times M}$ are the $N\times N$ identity matrix, $N\times M$ zero, and $N\times M$ one matrix.

## II. SYSTEM MODEL

Consider a single-user scenario consisting of a transmitter and a receiver communicating over a wireless channel. Let data symbols to be transmitted within one DFT-s-OFDM symbol be the elements of vector $\mathbf{d} \in \mathbb{C}^{N_d \times 1}$, where $N_d$ is the number of data symbols. In basic DFT-s-OFDM, first, the data symbols are mapped to the input of a DFT matrix denoted by $\mathbf{D} \in \mathbb{C}^{M \times M}$ via a mapping matrix $\mathbf{M}_t \in \mathbb{C}^{M \times M}$, where $M$ is the DFT size and $M = N_d$ as a special case. The output of the DFT is then mapped to a set of subcarriers in frequency domain via another mapping matrix $\mathbf{M}_f \in \mathbb{C}^{N \times M}$. Without loss of generality, the mapping matrix $\mathbf{M}_f$ can be constructed such that it allocates $M$ localized or interleaved subcarriers to achieve low PAPR. Finally, the output of the matrix $\mathbf{M}_f$ is converted to time domain via $\mathbf{F}^{\mathrm{H}}$ as

$$\mathbf{x} = \mathbf{F}^{\mathrm{H}} \mathbf{M}_f \mathbf{D} \mathbf{M}_t \mathbf{d}, \quad (1)$$

where $\mathbf{F}^{\mathrm{H}} \in \mathbb{C}^{N \times N}$ is the inverse DFT (IDFT) matrix and $N$ is the number of subcarriers. In this study, we denote the constrcution in (1) as plain DFT-s-OFDM [2, 8].

Let the channel impulse response (CIR) between the transmitter and the receiver be a vector $\mathbf{h} = [h_0\ h_1 \cdots h_{\mathcal{L}}]$ where $\mathcal{L} + 1$ is the number of taps. Assuming that the size of the cyclic prefix is larger than $\mathcal{L}$, the received signal vector $\mathbf{y}$ can be expressed as

$$\mathbf{y} = \mathbf{H}\mathbf{x} + \mathbf{n}, \quad (2)$$

where $\mathbf{H} \in \mathbb{C}^{N\times N}$ is the circular convolution matrix that models the interaction between the transmitted signal $\mathbf{x}$ and the channel $\mathbf{h}$, and $\mathbf{n} \in \mathbb{C}^{N\times N} \sim \mathcal{CN}(\mathbf{0}_{N\times 1}, \sigma^2 \mathbf{I}_N)$ is the additive white Gaussian noise (AWGN) with variance $\sigma^2$.

At the receiver, the operations applied at the transmitter are reversed by considering the impact of the multipath channel. The receiver operation can be expressed as

$$\tilde{\mathbf{d}} = \mathbf{M}_t^{\mathrm{H}} \mathbf{D}^{\mathrm{H}} \mathbf{Q} \mathbf{M}_f^{\mathrm{H}} \mathbf{F} \mathbf{y}, \quad (3)$$

where $\tilde{\mathbf{d}} \in \mathbb{C}^{N_d \times 1}$ is the estimated data symbol vector and $\mathbf{Q} \in \mathbb{C}^{M\times M}$ is the equalizer which removes the impact of the multipath channel. In this study, the equalizer $\mathbf{Q}$ is a diagonal matrix and derived by using the minimum mean square error (MMSE) criterion.

As can be seen in (1), data symbols are spread across frequency by the matrix $\mathbf{D}$ in DFT-s-OFDM. Therefore, the plain DFT-s-OFDM waveform does not support the introduction of frequency domain RSs in $M$-dimensional subspace spanned by $M$ columns of $\mathbf{F}^{\mathrm{H}}$. To allow the receiver to estimate the channel, the RSs can be transmitted with another DFT-s-OFDM symbol by using a fixed sequence, e.g., Zadoff-Chu sequence, as in LTE [2]. However, adopting two separate DFT-s-OFDM symbols reduces the data rate substantially as the number of estimated coefficients to extrapolate the channel frequency response may be significantly less than $M$. In order to insert RSs at some frequency tones, one may follow different strategies. One option is to puncture the information in the frequency domain by relying on the redundancy introduced by channel coding. However, this option does not yield recoverable DFT-s-OFDM at the receiver as the number of unknowns, i.e., $N_d(= M)$, is greater than the number of observations, i.e., $M - N_p$, within one symbol after the puncturing, i.e., $N_d = M > M - N_p$, where $N_p > 0$ is the number of punctured samples in frequency. In another method, one can reduce the number of data symbols as $N_d < M$ and change the size of $\mathbf{D}$ from $M$ to $N_d$ to accommodate the reference symbols within an $M$-dimensional subspace. However, reference symbols are generally not needed for all of the symbols in a frame or subframe. Thus, this option requires that both transmitter and receiver employ a DFT block with variable sizes, and may not be suitable for a radix-2 FFT implementation. In the third option, the number of data symbol $N_d \leq M$ is reduced while keeping the size of DFT as $M$ so that the number of unknowns is less than or equal to the number of observations after the puncturing, i.e., $N_d \leq M - N_p$. This option does not increase the transmitter complexity. However, the puncturing implicitly causes interference to the data symbols and it is not straightforward to recover the data symbols with a *low-complexity* receiver. In the following part of this paper, we will address this challenge and show that the data symbols can be recovered with a low-complexity receiver by employing a certain puncturing pattern and inserting zeros to specific locations before DFT-spread block at the transmitter.

## III. RECOVERABLE DFT-s-OFDM WITH FREQUENCY DOMAIN REFERENCE SYMBOLS

In this scheme, we introduce $N_z = M - N_d \geq N_p$ null symbols before DFT spreading as mentioned earlier and shown in Fig. 1. This allows for the number of observations to be greater than or equal to the number of unknowns after puncturing $N_p$ samples in frequency. We express the puncturing operation with the matrix $\mathbf{P} \in \mathbb{R}^{(M-N_p) \times M}$ and consider that $\mathbf{P}$ punctures one symbol every other $N_i$ symbols at the output of the DFT with an offset $S$, as shown in Fig. 1. Due to its periodic structure, the matrix $\mathbf{P}$ can be expressed as

$$\mathbf{P} = \mathbf{I}_{N_p} \otimes \begin{bmatrix} \mathbf{I}_S & & \mathbf{0}_{S \times N_i - S} \\ \mathbf{0}_{N_i - S \times S} & \mathbf{0}_{N_i \times 1} & \mathbf{I}_{N_i - S} \end{bmatrix}, \quad (4)$$

where $N_p = \frac{M}{N_i + 1}$ and $N_i + 1$ is integer multiple of $M$. Without loss of generality, the punctured vector is mapped to another vector in $M$-dimensional space by inserting $N_p$ zeros via a nulling matrix $\mathbf{N} \in \mathbb{C}^{M \times M - N_p}$ to accommodate frequency domain reference symbols denoted by $c_l$ where $l = 1, 2, \dots, N_p$. The reference symbols can be distributed uniformly in frequency to improve channel estimation performance at the receiver as shown in Fig. 1 [2]. In this case, one may choose the matrix $\mathbf{N}$ as

$$\mathbf{N} = \mathbf{I}_{N_p} \otimes [\mathbf{I}_{N_i} \quad \mathbf{0}_{N_i \times 1}]^T. \quad (5)$$

The overall transmit operation for punctured DFT-s-OFDM symbols can finally be expressed as

$$\mathbf{x} = \alpha \mathbf{F}^H \mathbf{M}_f \mathbf{NPDM}_t \begin{bmatrix} \mathbf{d} \\ \mathbf{0}_{N_z \times 1} \end{bmatrix}. \quad (6)$$

where $\alpha = \sqrt{\frac{N_d}{N_d - N_p}}$ is scalar which scales the energy of $\mathbf{x}$ to be $N_d$ again after the puncturing.

As discussed in Section II, the puncturing operation distorts the output of the DFT-spread implicitly and causes significant interference on the data symbols. The interference on the data and null symbols can be expressed as

$$\mathbf{r} = \mathbf{D}^H \mathbf{P}^H \mathbf{P} \mathbf{D} \mathbf{d}_e - \mathbf{d}_e, \quad (7)$$

where $\mathbf{d}_e \in \mathbb{C}^{M \times 1}$ is the mapped data symbols and can be obtained as $\mathbf{d}_e = \mathbf{M}_t [\mathbf{d}^H \quad \mathbf{0}^H_{N_z \times 1}]^H$ and $\mathbf{r} \in \mathbb{C}^{M \times 1}$ is the interference vector. The interference vector is not arbitrary as every other $N_I$ output of the DFT-spread block is nulled. By using the following lemma, we show the structure of the interference vector $\mathbf{r}$:

**Lemma 1 (Periodic Interference):** *Let $(X_n)$ be a sequence of size $M \in \mathbb{Z}$ for $n = 0, 1, \dots, M - 1$ and let $(Y_n)$ be another sequence obtained by zeroing every other $N_i$, $N_i \in \mathbb{N}$ elements of $(X_n)$ with an offset of $S$, $S \leq N_i$, $S \in \mathbb{N}_0$. Then, it is possible to decompose the IDFT of $Y_n$ as*

$$y_k = x_k + r_k, \quad \text{for } n = 0, \dots, M - 1, \quad (8)$$

*where $(y_k)$ is the IDFT of $(Y_n)$, $(x_k)$ is the IDFT of $(X_n)$, and $(r_k)$ is a sequence of size $M$ given by $r_k = p_k e^{j2\pi k \frac{S}{M}}$, for $k = 0, \dots, M - 1$ where $(p_k)$ is a period sequence with the period of $\frac{M}{N_i + 1}$.*

*Proof:* The elements of the sequence $(Y_n)$ can be expressed by using an auxiliary sequence $(R_n)$ as

$$Y_n = X_n + R_n, \quad (9)$$

where

$$R_n \triangleq \begin{cases} -X_n & \frac{n - S}{N_i + 1} \in \mathbb{Z} \\ 0 & \text{otherwise} \end{cases}. \quad (10)$$

Since the IDFT operation is linear, the IDFT of $(Y_n)$ can be expressed as $(y_k) = (x_k) + (r_k)$, where $(r_k)$ is the IDFT of $(R_n)$. The elements of $(r_k)$ can be calculated as

$$r_k = \sum_{n=0}^{M-1} R_n e^{j2\pi k \frac{n}{M}} \stackrel{(a)}{=} \sum_{m=0}^{\frac{M}{N_i+1}-1} -X_{(N_i+1)m+S} e^{j2\pi k \frac{(N_i+1)m+S}{M}}$$

$$\stackrel{(b)}{=} \underbrace{s_{k \bmod \left(\frac{M}{N_i+1}\right)}}_{p_k} e^{j2\pi k \frac{S}{M}} \quad (11)$$

where $(s_m)$ is the IDFT of $(-X_{(N_I+1)m+S})$ for $m = 0, \dots, \frac{M}{N_i+1} - 1$. In (11), (a) is true because $r_n$ is zero when $\frac{n-S}{N_i+1}$ is not an integer and (b) is true due to the periodicity of the exponential function $e^{-j2\pi k \frac{(N_i+1)m}{M}}$, which results in $p_k = p_{k + \frac{M}{N_i+1}}$. ∎

Lemma 1 has two important results. First, by using Lemma 1, one can deduce that the $k$th element of the vector $\mathbf{r}$ is $r_k = p_k e^{j2\pi k \frac{S}{M}}$ such that $p_k = p_{k+N_p}$. Secondly, it shows that the degrees of freedom for the interference vector $\mathbf{r}$ is $N_p$ as $p_k = $

$p_{k+N_p}$. Hence, one can regenerate the vector $\mathbf{r}$ by observing *only* $N_p$ elements of $\mathbf{r}$ that correspond to the *samples within one period of $p_k$* and inferring the rest of the vector $\mathbf{r}$ by using the relation of $p_k = p_{k+N_p}$. As a result, $\mathbf{M}_t$ should be chosen such that the location of the null symbols captures the samples at least for one period of $p_k$. Otherwise, the rank of $\mathbf{NPDM}_t$ becomes less than $N_d$ which would cause degradation at the receiver. Hence, Lemma 1 enlightens where to insert null symbols to allow the receiver to recover the data symbol without any distortion.

For example, let $M = 8$, $S = 0$, and $N_p = 2$, and assume that one chooses the input of the DFT block to be $(d_1, d_2, \ldots, d_6, 0, 0)$ (i.e., $N_z = 2$, $\mathbf{M}_t = \mathbf{I}_8$). Let $(x_1, x_2, \ldots, x_8)$ be the output of DFT. After discarding every $N_i = 3$ DFT outputs and replacing them by $(c_1, c_2)$, one gets $\{c_1, x_2, x_3, x_4, c_2, x_6, x_7, x_8\}$, which will be fed to IDFT block to generate time domain signal. At the receiver side, there are only 6 samples related to data symbols at the output of IDFT block. By neglecting the impact of noise for the sake of clarity and by using Lemma 1, one can show that IDFT of the equalized vector $\mathbf{d}_e$ is $(d_1 + p_1, d_2 + p_2, d_3 + p_1, \ldots, d_5 + p_1, d_6 + p_2, p_1, p_2)$ where the last two samples reveal the interference vector $\mathbf{r}$ as $p_k = p_{k+2}$. On the other hand, the selection of the data vector as $(0, d_1, 0, d_2, \ldots, d_6)$ does not allow the receiver to regenerate $\mathbf{r}$ as first and third samples carry the same interference sample after the puncturing.

### A. Low-complexity Receiver

At the receiver side, up to the frequency domain de-mapping operation, i.e., $\mathbf{M}_f^H$, the signal process is the same for both the plain DFT-s-OFDM and the proposed scheme. Unlike plain DFT-s-OFDM, the subcarriers that carry the reference signals at the output of DFT can be used for channel estimation (CHEST) with the proposed scheme. By using the estimated channel, the data bearing subcarriers are first equalized via $\mathbf{Q} \in \mathbb{C}^{M-N_p \times M-N_p}$ and the symbols at the output of equalizer are then mapped to the input of IDFT via $\mathbf{P}^H$. The output of IDFT $\mathbf{D}^H$ can be expressed as

$$\tilde{\mathbf{d}}_e = \frac{1}{\alpha}\mathbf{D}^H\mathbf{P}^H\mathbf{Q}\mathbf{N}^H\mathbf{M}_f^H\mathbf{F}\mathbf{y}, \tag{12}$$

where $\tilde{\mathbf{d}}_e \in \mathbb{C}^{M \times 1}$ is the received vector which includes the impacts of noise, equalization, and puncturing. Considering the structure of the interference due to the puncturing, a simple way to recover the data symbols is

$$\tilde{\mathbf{d}} = \mathbf{M}_{t,d}^H \tilde{\mathbf{d}}_e - \mathbf{R}\mathbf{M}_{t,r}^H \tilde{\mathbf{d}}_e \tag{13}$$

where $\mathbf{M}_{t,d} \in \mathbb{C}^{M \times N_d}$ and $\mathbf{M}_{t,r} \in \mathbb{C}^{M \times N_z}$ as the submatrices of $\mathbf{M}_t$ as $\mathbf{M}_t = [\mathbf{M}_{t,d} \ \mathbf{M}_{t,r}]$, and $\mathbf{R} \in \mathbb{C}^{N_d \times N_z}$ is the reconstruction matrix calculates the distortion due to the puncturing based on the relation of $r_k = p_k e^{2\pi k \frac{S}{M}}$ and $p_k = p_{k+N_p}$ dictated by Lemma 1. As a special case, when $S = 0$ and $\mathbf{M}_t = \mathbf{I}_M$, $\mathbf{R}$ becomes a simple repetition matrix given by

$$\mathbf{R} = \mathbf{1}_{N_l \times 1} \otimes \mathbf{I}_{N_z}, \tag{14}$$

which simplifies the receiver structure substantially. For example, if $\mathbf{d}_e$ is $(d_1 + p_1, d_2 + p_2, d_3 + p_1, \ldots, d_5 + p_1, d_6 + p_2, p_1, p_2)$, $\mathbf{R}$ replicates the last two samples by $N_l = 3$ times and the receiver can recover the data symbols by subtracting the replicated vector from the rest of the samples of $\tilde{\mathbf{d}}_e$ as expressed in (13). The proposed scheme in this case introduces only $N_d$ extra complex addition to the plain DFT-s-OFDM receiver.

### B. Iterative Receiver

Although the method discussed in Section III.A allows a low-complexity receiver, it enhances the noise as two noisy observations are added by (13). One effective way of mitigating the noise enhancement is to use an iterative receiver which aims at removing the noise on the second part of (13), i.e., distortion due to the puncturing. To this end, for the $i$th iteration, the data symbols are estimated by

$$\tilde{\mathbf{d}}^{(i)} = \mathbf{M}_{t,d}^H \tilde{\mathbf{d}}_e - \mathbf{R}\mathbf{M}_{t,r}^H \tilde{\mathbf{d}}_e^{(i-1)}, \tag{15}$$

where $\tilde{\mathbf{d}}_e^{(0)} = \tilde{\mathbf{d}}_e$. The estimated data symbols $\tilde{\mathbf{d}}^{(i)}$ are then mapped to closest symbol in the constellation by a non-linear function $f(\cdot)$, i.e., demodulation, and $\tilde{\mathbf{d}}_e^{(i+1)}$ is prepared for the next iteration as

$$\tilde{\mathbf{d}}_e^{(i+1)} = \mathbf{D}^H \mathbf{P}^H \mathbf{P} \mathbf{D} \mathbf{M}_t \begin{bmatrix} f(\tilde{\mathbf{d}}^{(i)}) \\ \mathbf{0}_{N_l \times 1} \end{bmatrix}. \tag{16}$$

Since $\tilde{\mathbf{d}}_e^{(i+1)}$ is generated after the decision is made by $f(\tilde{\mathbf{d}}^{(i)})$, it removes the noise from the second part of (15) effectively and leads to a better estimate of $\tilde{\mathbf{d}}$ for the $(i + 1)$th iteration. Compared to the receiver structure given in Section III.A, (15) and (16) increase the receiver complexity as it involves multiple DFT operations and demodulation operations.

### C. Generalization and Constraints

It is important to emphasize that the proposed scheme introduces some conditions on the puncturing pattern, the number of reference signals $N_p$, the number of null symbols $N_z$, and the pattern of the null symbols. First, the receive structures discussed in Section III.A and Section III.B exploit the fact that every $N_i$ other output of the DFT with an offset $S$ are punctured. Second, $N_z \geq N_p$ must hold and the pattern of $N_z$ null symbols at input of DFT-spread block should capture at least one period of distortion due to the puncturing to yield a recoverable a DFT-s-OFDM symbol. One simple way of doing this would be to consider $N_z$ adjacent null symbols.

There is also room to increase the performance of the receiver performance. For example, one simple way of improving the receiver performance is given in Section III.A where the number of null symbols are increased to more than the number of punctured symbols, i.e., $N_z > N_p$. In this case, the receiver can combine the samples to calculate a more reliable interference vector at the expense of a less spectrally efficient scheme. The receiver structure given in Section III.B is also improved by including the channel coding decoder along with demodulation on the feedback branch [7].

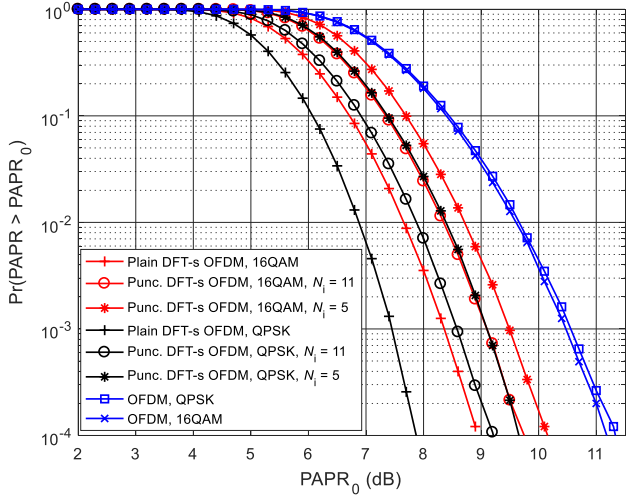

Fig. 2. PAPR performance.

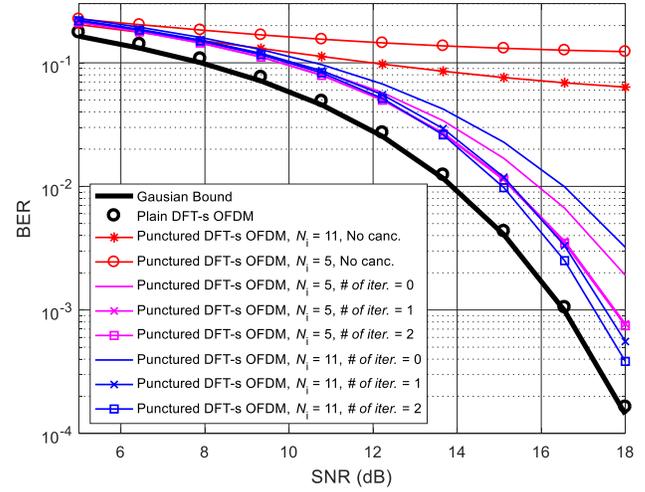

Fig. 3. BER performance in AWGN channel (16QAM).

Without loss of generality, the proposed scheme can be expended to multiple DFT blocks [4]. In addition, if the subcarriers that are discarded at the transmitter side are not replaced by RSs (i.e., replaced by zeros), the corresponding subcarriers at the receiver DFT output can also be used for noise or interference power estimation.

## IV. NUMERICAL RESULTS

In this section, we demonstrate PAPR and uncoded BER performance of the proposed scheme through simulations based on a LTE [2]. We consider an uplink scenario where the IDFT size $N$ is 2048 subcarriers and the sampling rate of the radio is 30.72 MHz (i.e., corresponds to the subcarrier spacing of 15 kHz for an OFDM symbol). In the simulations, $M = 48$ subcarriers are allocated for uplink DFT-s-OFDM transmission. To evaluate the proposed scheme, we consider 16QAM, $N_i \in \{5,10\}$, and $S = 0$. We employ the receiver structures discussed in Section III.A and Section III.B. For multipath channel, we adopt Extended Pedestrian A (EPA) and Extended Vehicular A (EVA) channel models unless otherwise stated. The RSs are generated based on a cyclically extended Zadoff-Chu sequence.

### A. PAPR Performance

In Fig. 2, we compare the PAPR performance of punctured DFT-s-OFDM, plain DFT-s-OFDM, and OFDM. Among the waveforms, plain DFT-s-OFDM has the best PAPR results due to its single carrier nature whereas OFDM has the worst PAPR results as it has simultaneously active $M = 48$ subcarriers. DFT-s-OFDM loses its single carrier structure after the output of DFT-spread block is punctured and replaced by a reference signal. While the former impact increases the variation of the data symbols (as additional distortion is added to the puncturing), the latter increases the PAPR by adding another parallel signal operating at different frequency. Nevertheless, since the PAPR of ZC sequence is very low [2], the PAPR of the total signal is still lower than OFDM. For $N_i = 5$ and $N_i = 11$, there are $N_p = 8$ and $N_p = 4$ punctured samples in frequency, respectively. Hence, the case where $N_i = 5$ introduces more PAPR as compared to the case where $N_i = 11$.

### B. BER Performance in AWGN

In Fig. 3, we evaluated the BER performance of the aforementioned schemes in AWGN. The BER performance of the plain DFT-s-OFDM is aligned with the Gaussian BER bound. However, when the puncturing operation is introduced to the DFT-s-OFDM, the performance of plain DFT-s-OFDM receiver dramatically decreases as shown in Fig. 3 for the cases of $N_i = 5$ and $N_i = 11$ (i.e., the plain DFT-s-OFDM receiver ignores the puncturing and no cancellation mentioned in Section III.A is applied). However, the receiver structures which exploit the periodic interference due to the puncturing significantly help the BER performance and the error floor is removed. When the number of iterations is set to zero, the receiver structure corresponds to the one described in Section III.A. In this case, the punctured DFT-s-OFDM is approximately 2 dB off from the Gaussian BER bound as two noisy components are added[1]. However, when the number of iterations is more than 0, the interference due to the puncturing exhibits a better estimation using decision feedback. In this case, the BER performance is approximately 1 dB off from the Gaussian BER bound. The simulation results indicate that there is no notable improvement after 1-2 iterations.

---
[1] The noise enhancement is less than 3 dB as the scalar $\alpha > 1$ in (6) normalizes the symbol energy after the puncturing.

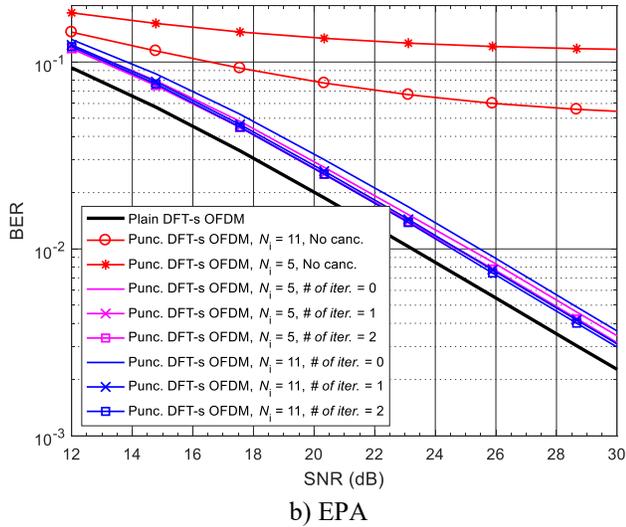

b) EPA

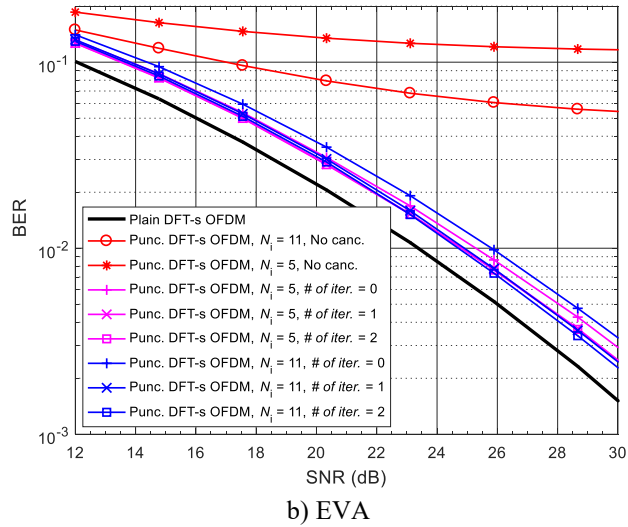

b) EVA

Fig. 4. BER performance in fading channel (16QAM).

## C. BER Performance in Fading Channel

In Fig. 4a and Fig. 4b, we compare the BER performance in two different multipath fading channel models, i.e., EPA and EVA, respectively. To compare the schemes, we assume that channel estimation is ideal. In both cases, the results indicate that the interference due to the puncturing significantly deteriorates the BER performance of the plain DFT-s-OFDM receiver (i.e., the BER curves with the label of no cancellation). Since the number of punctured samples in the case where $N_i = 5$ are more than that of $N_i = 11$, more degradation is observed in the case of $N_i = 5$. However, when the receivers that exploit the periodic structure of interference are employed, the error floor is removed from both cases. Without any iteration, approximately 2 dB worse BER performance is observed as compared to plain DFT-s-OFDM. With iterations, the BER performance is improved by 1 dB approximately for the case where $N_i = 5$ and $N_i = 11$. Similar to the results in AWGN channel, the receiver described in Section III.B. settles after 1-2 iterations.

## V. CONCLUDING REMARKS

In this study, we propose a new scheme based on DFT-s-OFDM, in which the DFT-s-OFDM symbol contains RSs in the frequency domain. The proposed scheme can be beneficial to achieve a framework in which the RS allocation in time and frequency is common for DFT-s-OFDM and OFDM-based waveforms. The proposed scheme is realized by puncturing the output of the DFT-spread block and replacing the punctured samples with RSs. We prove that puncturing the interleaved samples at the output of a DFT operation causes a periodic interference to data symbols at the input of DFT. By exploiting this fact, we show that inserting zeros to certain locations before DFT-spread operation at the transmitter allows the receiver to compensate for the interference due to the puncturing with low-complexity operations. Simulation results show that punctured DFT-s-OFDM with frequency domain RSs has lower PAPR than OFDM and SNR loss can be as low as 1 dB in AWGN and 1.5 dB for fading channel with the investigated receiver structures as compared to plain DFT-s-OFDM.